\documentclass{PoS}

\title{F(750), We Miss You as a Bound State of 6 Top
and 6 Antitop Quarks, Multiple Point Principle}

\ShortTitle{F(750), We Miss You!}

\author{\speaker{Holger F. Bech Nielsen}
\thanks{hbech@nbi.dk, hbechnbi@gmail.com}
\\
        Niels Bohr Institute, Blegdamvej 15-21,
DK 2100 Copenhagen {\O}, Denmark

{\bf D.L.~Bennett}\\
Brooks Institute, Copenhagen, Denmark\\
{\bf C.R.~Das}${}$
\footnote{crdas.crdas@gmail.com,
          crdas@cftp.ist.utl.pt,
          das@theor.jinr.ru}\\
BLTP, JINR, Dubna, Moscow, Russia\\
{\bf C.D.~Froggatt}${}$
\footnote{c.froggatt@physics.gla.ac.uk}\\
Glasgow University, Glasgow, UK\\
{\bf L.V.~Laperashvili}${}$
\footnote{laper@itep.ru}\\
ITEP, National Research Center ``Kurchatov Institute'', Moscow, Russia
}

\abstract{We review our
speculation, that {\em in the pure
Standard Model} the exchange of Higgses, including
also the ones ``eaten by $W^{\pm}$ and Z'', and of
gluons together make a {\em bound state of 6 top plus
6 anti top quarks bind} so {\em strongly} that its mass gets
down to about 1/3 of the mass of the collective
mass 12 $m_t$ of the 12 constituent quarks. The true importance
of this speculated bound state is that it makes it
possible to uphold, even inside the Standard Mode, our
proposal for what is really a new law of nature saying
that there are {\em several} phases of empty space,
{\em vacua},
all having very {\em small energy densities} (of the
order of the present energy density in the universe).
The reason suggested for believing in this new law called
the ``Multiple (Criticality) Point Principle'' is, that
estimating the mass of the speculated bound state using
the ``Multiple Point Principle'' leads to two consistent
mass-values; and they even agree with a crude bag-model
like estimate of the mass of this bound state. Very
unfortunately the statistical fluctuation so popular last
year when interpreted as
the digamma resonance F(750) turned out not to be a real resonance,
because our estimated bound state mass is just around the mass of
750 GeV. }

\FullConference{Corfu Summer Institute 2016
"School and Workshops on Elementary Particle Physics
and Gravity"\\
		31 August - 23 September, 2016\\
		Corfu, Greece}

\begin{document}

\section{Introduction}
{\bf Main ideas}
The main point of the present contribution is to
attempt to make the listeners believe in our proposal
for a ``new law of nature'' - which we shall call the
``Multiple (criticality) Point Principle''\cite{5mp,6mp,7mp,
8mp,9mp,DonThesis,10mp,11mp,12mp,13mp,14mp,15mp,16mp,17mp, Kawana1,
%Kawana2,
Kawana3, Kawana4, Kawana5,Yamaguchi,KO}, shortened to
MPP - which delivers some information about
the coupling constants (or better the parameters of
the theory, say e.g. the Standard Model). In realizing
the picture, which we suggest in connection with this new
law of nature, we further need to have that there exists
an extremely strongly bound state\cite{1nbs, 2nbs, 3nbs,
4nbs, 5nbs,6nbs,7nbs, 8nbs,9nbs,10nbs,11nbs, 12nbs,13nbs,14nbs,
LNvacuumstability,mass} of six top plus six
antitop quarks, and thus our main points are:

\begin{itemize}
\item{\bf Strongly Bound State}\\
We speculate that mainly due to exchange
of Higgs bosons a system of 6 top plus
6 anti top quarks bind so strongly as
to make a {\bf bound state} with appreciably
lower mass than that of 12 separate
quarks and anti quarks.
\item{\bf Multiple Point Principle}\\
We propose a {\bf new law of nature (MPP)}
 saying,
 that - somewhat mysteriously may be -
the {\bf coupling constants} and other
parameters, such as the Higgs mass squared,
{\bf get adjusted} so as to guarantee, that there are
{\em several vacua all with very small
energy
densities (=cosmological constants).}
\end{itemize}

By a strange accident there appeared around Christmas
2015 a peak in the spectrum, now known to be statistical
fluctuation, but at that time  looking
like a resonance produced in LHC and decaying into pairs
of photons with a mass of 750 GeV\cite{1,2,3,killingICHEP}
- thus called a digamma
F(750) -. This was a strange accident, because it is
the main point of the present article, that we calculate
the mass of the bound state of the 6 top + 6 anti tops
in our scheme - speculation inside pure Standard Model  -
in three different ways using our ``law'' MPP, and obtain
within our very low accuracy something perfectly consistent
with the mass being 750 GeV! We therefore really miss
this by now no longer statistically viable digamma state.

\begin{figure}
\centering
\includegraphics[scale=0.5]{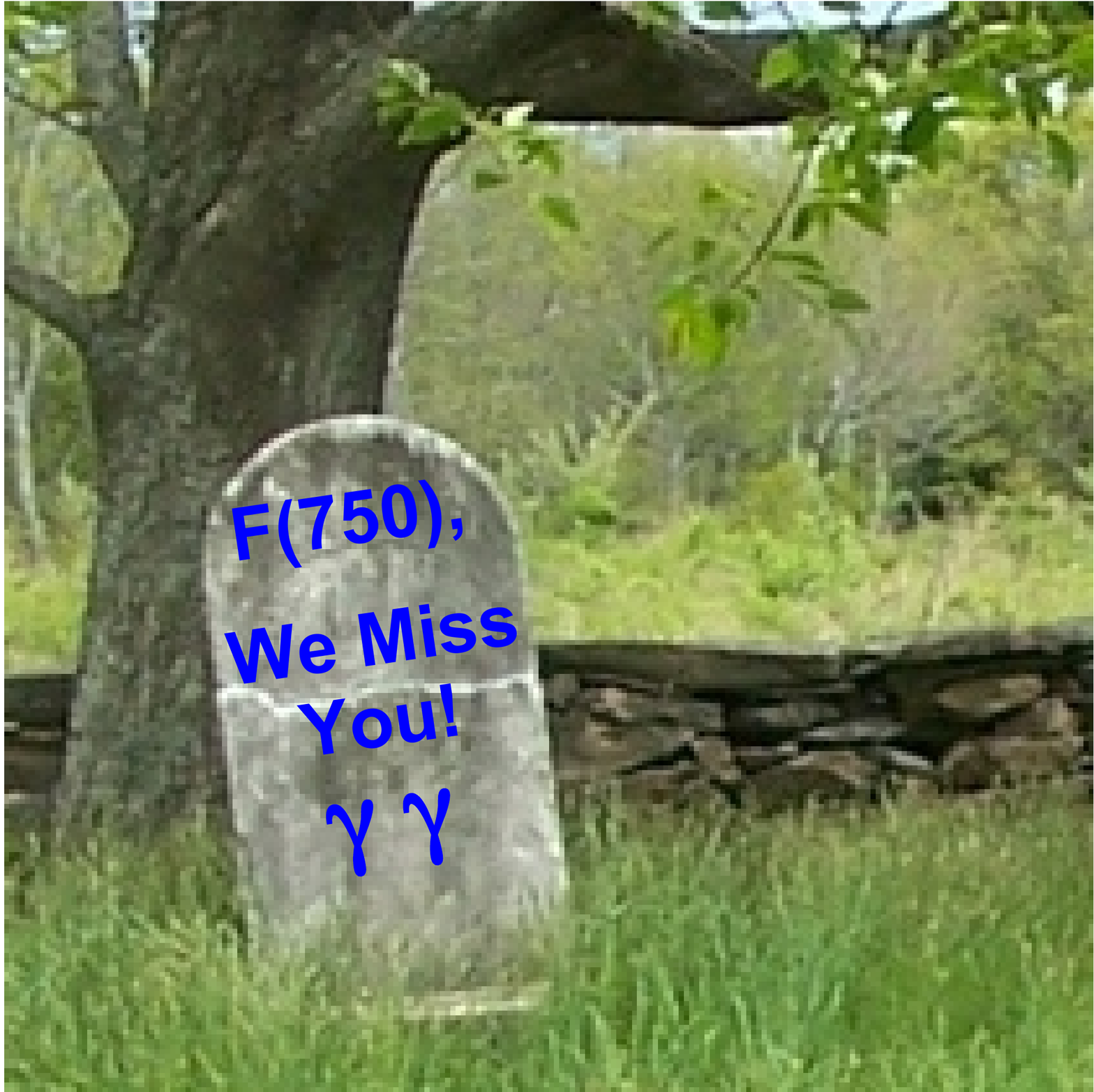}
\caption{We miss you!}
\end{figure}

An extremely tiny hope may though today be represented by the fact
that a new presumably again purely statistical fluctuation
at LHC corresponds to producing a resonance with mass 650 GeV decaying into
two Z-bosons. The experimental mass determinations are
so that 650 GeV is quite distinguishable from 750 GeV, but
our theoretical accuracy in predicting the mass of our bound
state is not sufficient to distinguish 650 GeV from 750 GeV,
and so for us the new 650 GeV fluctuation would be fine.
%for us.
According to our estimates \cite{FN750} the easiness of
observing
the decay of the 6t + 6 $\bar{t}$ bound state decaying into $\gamma\gamma$
and into $ZZ$ are rather similar, although the dominant decay
would be into $t\bar{t}$, but the latter may be more difficult
``see''.

\begin{figure}
\centering
\includegraphics[scale=0.5]{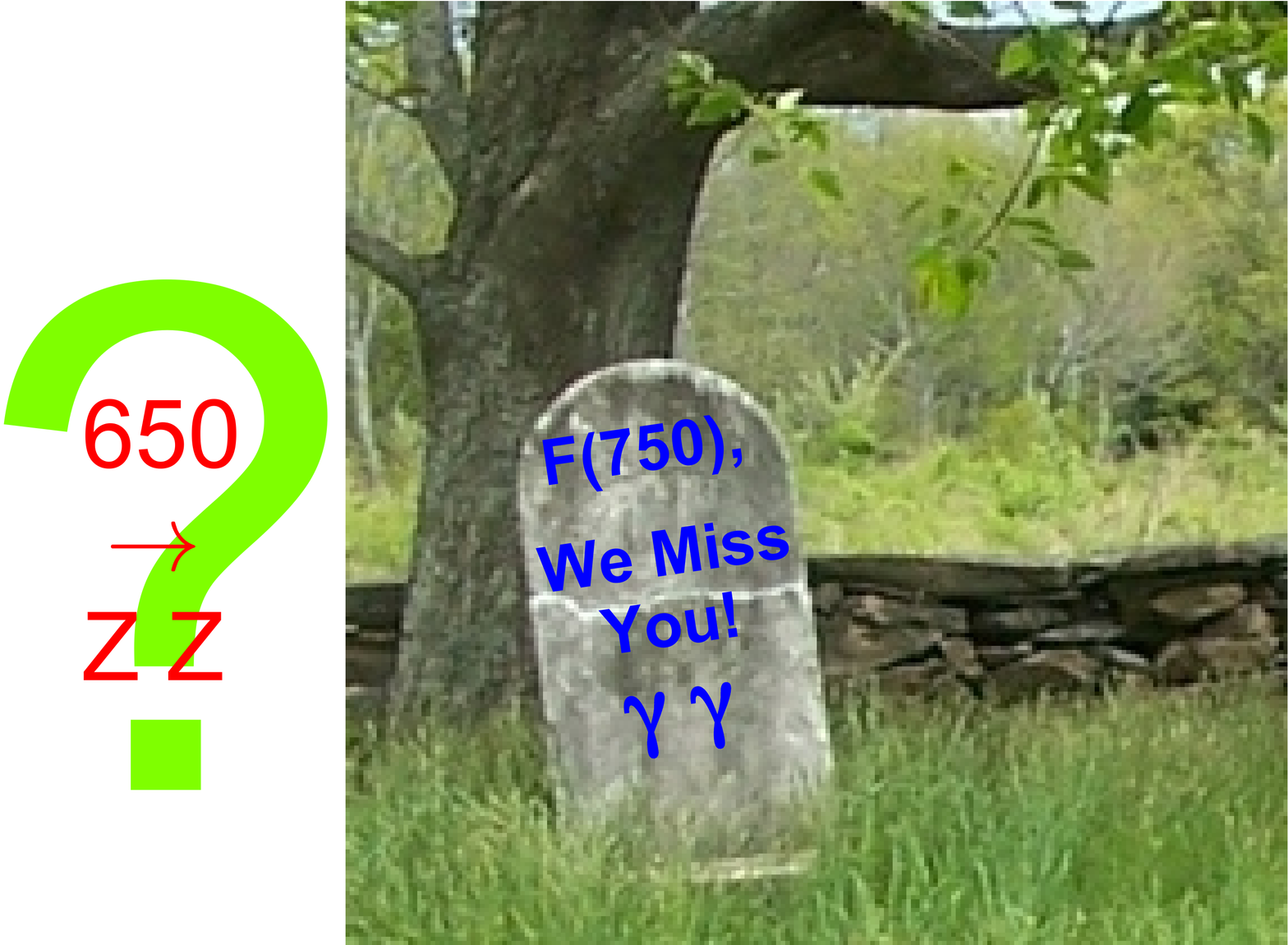}
\caption{We miss you! Must have a reason!}
\end{figure}

\section{Multiple Point Principle (MPP)}
Our proposal for a new law of nature
- Multiple Point Principle (MPP) -
(first by Don Bennett and myself) means
that {\bf there shall exist several vacua with
very small energy density.}

{\bf Three Vacua in Standard Model:}
For simplicity and trustability we in this
talk restrict ourselves to the {\bf pure
Standard Model}
and only the following {\bf three} vacua:

\begin{itemize}
\item{{\color{blue}Present}}: The vacuum, in which we
live, in the sense that, if we in
practice
find a place with zero  density of
material, then that region is in the state
of the ``present vacuum''.
\item{{\color{blue}High (Higgs) field
vacuum}}:
This vacuum is a state, in which the Higgs
field is at a minimum in the
Higgs-effective potential $V_{eff}(\phi_H)$
having a value
of the Higgs field near $\phi_H
\sim 10^{18}$ GeV.
It is known that, in the pure Standard Model,
it seems that the energy density of this
vacuum is slightly negative (with 3
standard deviations from being just zero).
\item{{\color{blue}Condensate vacuum}}:
This third
vacuum is a very speculative possible
state inside the pure Standard Model,
which contains a lot of strongly bound
states distributed smoothly in space, each bound
from 6 top + 6 anti
top quarks.
\end{itemize}

{\bf Can use Multiple Point Principle
together with Any Model (side-remark)!}

In the present talk I shall concentrate
on the version ``several vacua, that all
have very {\bf small energy densities}'' i.e. MPP.

But an older  version had it:
``Several vacua all have the {\bf same}
{\bf energy density} (with some accuracy,
that
can be discussed)'' in MPP.

Also one does not need to  assume just the
Standard Model as I shall do in the
present talk. For instance  Roman
Nevzorov with some of us (Froggatt and me)
\cite{15mp,16mp,17mp,Roman1,Roman2,
Roman4,Roman6,Roman7,Roman8,Roman9,Roman10,Roman11,
Roman12,Roman13,Roman15,Roman16,Roman17,Roman18,
Roman19,Roman20,Roman21}.
assumed a supersymmetry only broken tinily
in one vacuum, but much more strongly broken in
e.g. the present vacuum.

The works with Roman Nevzorov  extend the
 Standard Model with Susy etc. \cite{15mp,16mp,17mp,Roman1,Roman2,
Roman4,Roman6,Roman7,Roman8,Roman9,Roman10,Roman11,
Roman12,Roman13,Roman15,Roman16,Roman17,Roman18,
Roman19,Roman20,Roman21}.
(Otherwise in the present talk I only keep the Standard
Model.)

Assuming some of the vacua to have
only tinily broken susy, a
tiny {\bf cosmological constant} in the
almost
susy unbroken vacuum could be {\bf transferred
- by means of} a very accurate {\bf  MPP -} to the present
vacuum, and a rather
successful fitting/derivation of the
astronomically determined cosmological
constant could be achieved! (An extra
five-plet could make the fitting extremely good.)

Also the original idea on the basis of which we - Don
Bennett and I -  invented the
``multiple point principle'' was based
on a model called AntiGUT\cite{AGUTold1,
AGUTold2, AGUTold3, AntiGUT}, which extends
the Standard Model. However it has first  new physics
rather close to the Planck scale, actually
in it  each family of fermions has
its own system of gauge groups. The
gauge bosons are also in families!

{\bf Finetuning of Parameters, Couplings}:

Our ``multiple point principle'' is really
just an assumption about the coupling
constants - in the Standard Model, if
we as in this talk take the model to be
pure Standard Model - being
{\bf finetuned} so as to make the three
vacua proposed have just zero energy
density $V_{present}, V_{condensate},
V_{high \; field}=0$ (with say the
accuracy of the order
of the astronomically found energy density
in the ``present vacuum'' $\sim$ 75 \% of
the total energy density in the present
universe).

{\bf I.e. MPP provides 3 restrictions
between the parameters of the model in
question, here the Standard Model, from
which the $\sim$ zero energy densities
in all
three vacua follows.}

{\bf Multiple Point Principle means
Relations between the couplings and
other parameters:}

\begin{eqnarray}
V_{present}(\Lambda_{CC},g_t, m_H^2,
\Lambda_{QCD}, ...)
&=&0\\
V_{condensate}(\Lambda_{CC},g_t, m_H^2,
\Lambda_{QCD}, ...)
&=&0\\
V_{high \; field}(\Lambda_{CC},g_t, m_H^2,
\Lambda_{QCD},
...)&=&0
\end{eqnarray}
Here we wrote explicitly the following
parameters of the Standard Model:
\begin{eqnarray}
\Lambda_{CC} &:& \hbox{The cosmological constant}\\
g_t &:& \hbox{The top Yukawa coupling}\\
m_H^2&:& \hbox{Higgs mass squared}\\
\Lambda_{QCD} &:& \hbox{The scale parameter
of QCD}
\end{eqnarray}

Whether these parameters are renormalized
or bare does not matter so much here.

$V_{present}, V_{cindensate}, V_{high \; field}$
are  the vacuum energy densities for
the three speculated vacua.

{\bf Use of Multiple Point Principle:}

Taking the experimental values for all the
Standard Model parameters except for
say $\Lambda_{CC}, m_H^2$ and
$g_t$, we could look at it such that e.g.
$V_{present} =0$ fixes the cosmological
constant $\Lambda_{CC}$ to essentially
zero. (It is very small indeed.) Then
$V_{high \; field} =0$ (meaning the energy
density of the vacuum having the
very high Higgs field $\phi_H \approx
10^{18} GeV$) could be taken to
predict the Higgs mass, and the
$V_{condensate} =0$ to predict, say, the
$g_t$ Yukawa coupling. In fact Colin
Froggatt and
 H.B.Nielsen {\bf PRE}dicted the Higgs mass
many years ago\cite{9mp} to be $135$ GeV $\pm$ 10 GeV
from such an MPP-assumption.

In fact the speaker (H.B.N.) was portrayed long
before the Higgs was found experimentally at LHC
together with Mogens Lykketoft - a previous
member of the Danish cabinet - with the number
135 GeV $\pm$ 10 GeV
written partly behind the head of this member of the
cabinet, meaning our PREdiction for the Higgs boson
mass. Since the much later determined experimental
mass turned out to be $125.06$ GeV $\pm$ .21 GeV
(statistical uncertainty) $\pm$ 0.11 (systematic uncertainty),
we had agreement for our PREdiction deviating only by
one standard deviation (of {\em our} theoretical uncertainty
in the old time in the 90s):

\begin{figure}
\centering
\includegraphics[scale=2.0]{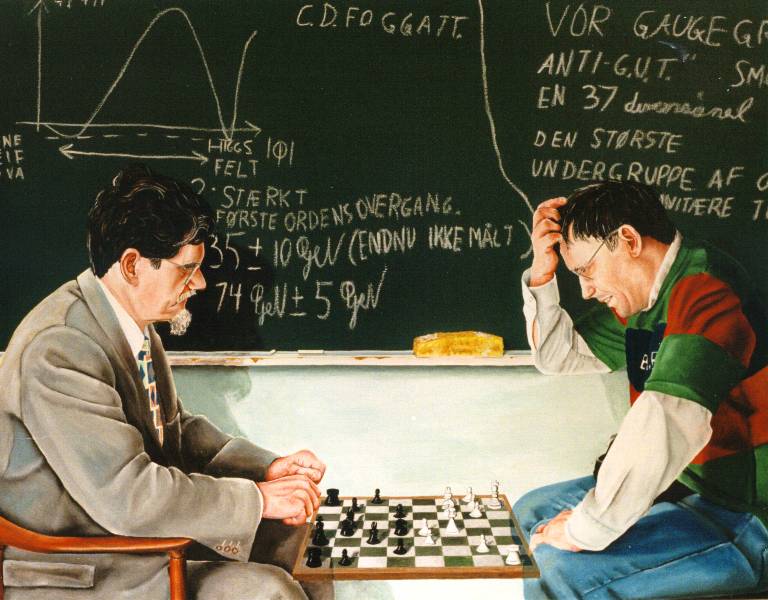}
\caption{Chess!}
\end{figure}

Having in mind that the ``present vacuum'' as well
as the ``high (Higgs) field vacuum'' basically
correspond to two different minima in the effective
potential $V_{eff}(\phi_H)$ for the Higgs field $\phi_H(x)$, we
see that the statement
of these two vacua having the same energy density means,
that there is in the effective potential $V_{eff}(\phi_H)$
as a function of the Higgs field strength $\phi_H$
two {\em equally deep minima}. Our PREdiction consisted
in fitting the Higgs mass $m_H$ to arrange the effective
potential for the Higgs field $V_{eff}(\phi_H)$ to have
such {\em two minima of equal depth}.

\begin{figure}
\centering
\includegraphics[scale=0.5]{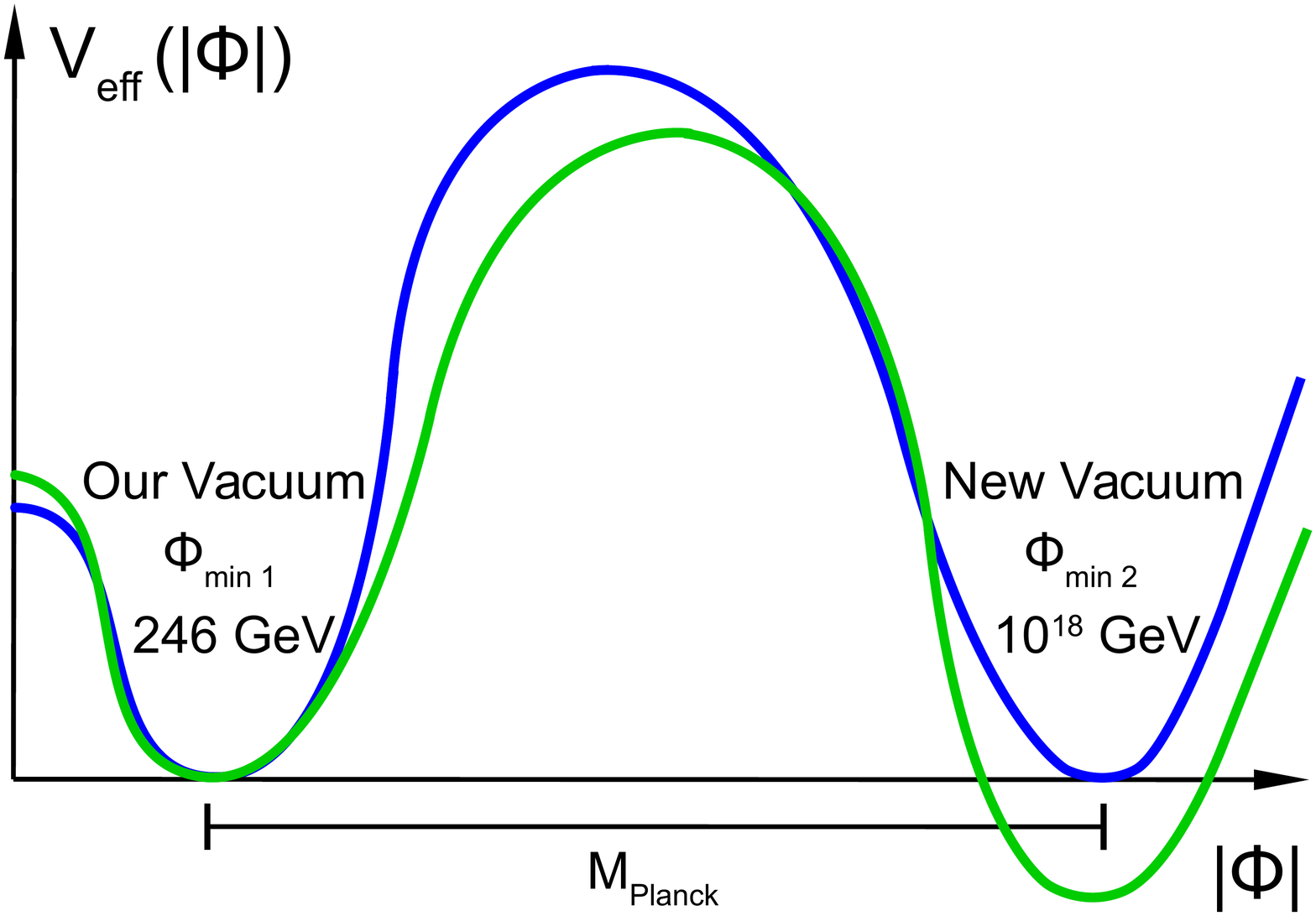}
\caption{Two different minima in the effective
potential for the Higgs field.}
\end{figure}

On the figure 4 the curve touching the abscissa axis in
two points is the one corresponding to the Higgs mass
as predicted from the ``Multiple Point Principle'', while
the curve crossing the abscissa a couple of times and
only touching it at the ``present vacuum'' Higgs field
of the well-known $246$ GeV corresponds to a somewhat
lower value of the Higgs mass. With the modern top mass value
and the very accurate theoretical calculations performed
in \cite{Deg,But}, the Higgs mass predicted from the ``multiple
point principle'' would be 129.4 GeV rather than the old
estimate 135 GeV. The uncertainty in the top-quark mass
measurement is indeed very important and it might be better
to look at the ``multiple point principle''-prediction
as giving a relation between the top and the Higgs masses.
In any case: while in the old accuracy of calculation
we got only one standard deviation from the by now measured
Higgs mass, the present deviation is rather 3 standard
deviations because of the increased top-quark mass measurement
accuracy,
in spite of that the predicted Higgs mass has indeed moved
in the good direction from the 135 GeV towards 129.4 GeV;
but the decrease in the uncertainty has been even faster.

It should also be mentioned that Colin Froggatt, Yasutaka Takanishi
and the speaker \cite{Yasutakamp} proposed the alternative idea of
rather than quite degenerate vacua (having same energy
density) to require just META-stability, which should
be expected as what should be important in as far as
that could influence whether our vacuum - the present
vacuum - decays to the high Higgs field or not. This
meta-stability - also long before the Higgs was found
experimentally at LHC - PREdited a mass of the Higgs
being 121 GeV. Actually this was even a bit closer
to the actual Higgs mass than our old 135 GeV. But the deviation
of the meta-stability PREdiction is to the opposite side:
Relative to ``META-multiple point principle'' the
experimental Higgs mass 125 GeV is slightly to the high
side of the PREdicted one.

{\bf A Remarkable Side-Point:}

The second minimum in the Higgs effective
potential corresponding to what, we call
the ``High field'' vacuum, has an
expectation value for the Higgs
field $\phi_H$, which is {\bf remarkably
close} (order of magnitudewise) {\bf to
the Planck energy scale}!
This should not be an accident, but rather
explained:
the Planck energy scale is the ``fundamental
physics scale'' for both energy and Higgs
fields; they have the same dimension.

Two of the vacua, which we discuss today
have  for some reason exceptionally
small,
 say,  Higgs field, while  the
``high field'' vacuum has the ``normal''
order of unity in Planck units value for
{\em its} Higgs expectation value.
So we rather ask the question:

{\color{blue} Why do the two vacua,
``present vacuum'' and
``condensate vacuum'', not have Planck
scale, say, Higgs fields? }
(Let me for the moment postpone a discussion of
the fact that
indeed we have an explanation\cite{6nbs,7nbs} from
the ``multiple point principle'', that
these two vacua  have
exceptionally small Higgs expectation
values scale.)
%\end{frame}

\section{Reason}
%\begin{frame}
{\bf But Why should we believe in the postulate of the Multiple
Point Principle ?}
\begin{itemize}
\item{{\bf Need Coupling Explanations}}:

It is clear that there are some parameters in the
Standard Model that take so special values, that it
cries for an explanation; there are fine tuning problems:
\begin{itemize}
\item Even with great effort e.g. Graham Ross\cite{Ross}
could not get the factor $\Delta$ with the inverse of
which the Higgs is too light further down
than about 1/20.
\item Cosmological
constant ? Why is it so small ?
\end{itemize}
\item{\bf Derivations}: In models which
allow somehow influence from the future\cite{Future} to
adjust coupling constants one may make
some ``derivations'' of MPP.
\item{\bf Empirical}: But really it is the main
point of today's talk to deliver some empirical support (our
PREdiction of the Higgs mass, and
two derivations of the same bound state mass)
\end{itemize}

{\bf 1. Theoretical Reason: ``Plural of Cosmological Constant''}

\begin{enumerate}
\item[1] We have to assume that the energy density in the
``present vacuum'' is very
small compared say to the Planck energy\cite{Weinberg}
density, because it has been well-known since
long before the measurements with
supernovae A1 settled it to be non-zero.
\item[2] This assumption does not become
essentially less beautiful or more
complicated by ``putting it in plural'':
{\bf
Several vacua have very small energy density/ cosmological
constant}
compared to say
the Planck energy density or the Higgs
energy density or compared to  most high energy physics
contributions.
\end{enumerate}

(This argument came from private conversation with L. Susskind.)

But each time you fix the energy of one
more vacuum energy density you get one
more relation between the parameters/
couplings of the theory.

It may even be more beautiful to talk
about several/all vacua than just a
special one that must be specified.

{\bf 2. Theoretical Reason: Extremizing Something,
Positive Energy}
\begin{enumerate}
\item[A.] Assume that energy-density
should be positive or zero. i.e. the bottom
in the Hamiltonian density is at least zero.
{\color{red} This will restrict the
coupling constants and other parameters
- e.g. Higgs mass - to some polyhedron-like figure with
curved sides, where the sides
correspond to one possible vacuum or
the other having just zero energy
density.}
\item[B.] Assume that the couplings and
parameters inside the positivity
restriction are selected by {\bf
minimizing something / some function
of these couplings and parameters} (a
generic or ``random'' function).

Using the {\em anthropic principle}\cite{anthropic} the function
could suggestively be the number of human
beings in the universe resulting with the
couplings etc. in the point in
parameter-space considered.
\end{enumerate}

On the figure 5 I have drawn the contour curves
- really contour surfaces - for the function to be
minimized.
%of codimension one -
%and marked them with numbers. The by
%positive allowed coupling parameter
%cobinations is symbolized by the
%{\color{violet} violet}
%polyhedron-like region with curved sides.

\begin{figure}
\centering
\includegraphics[scale=0.5]{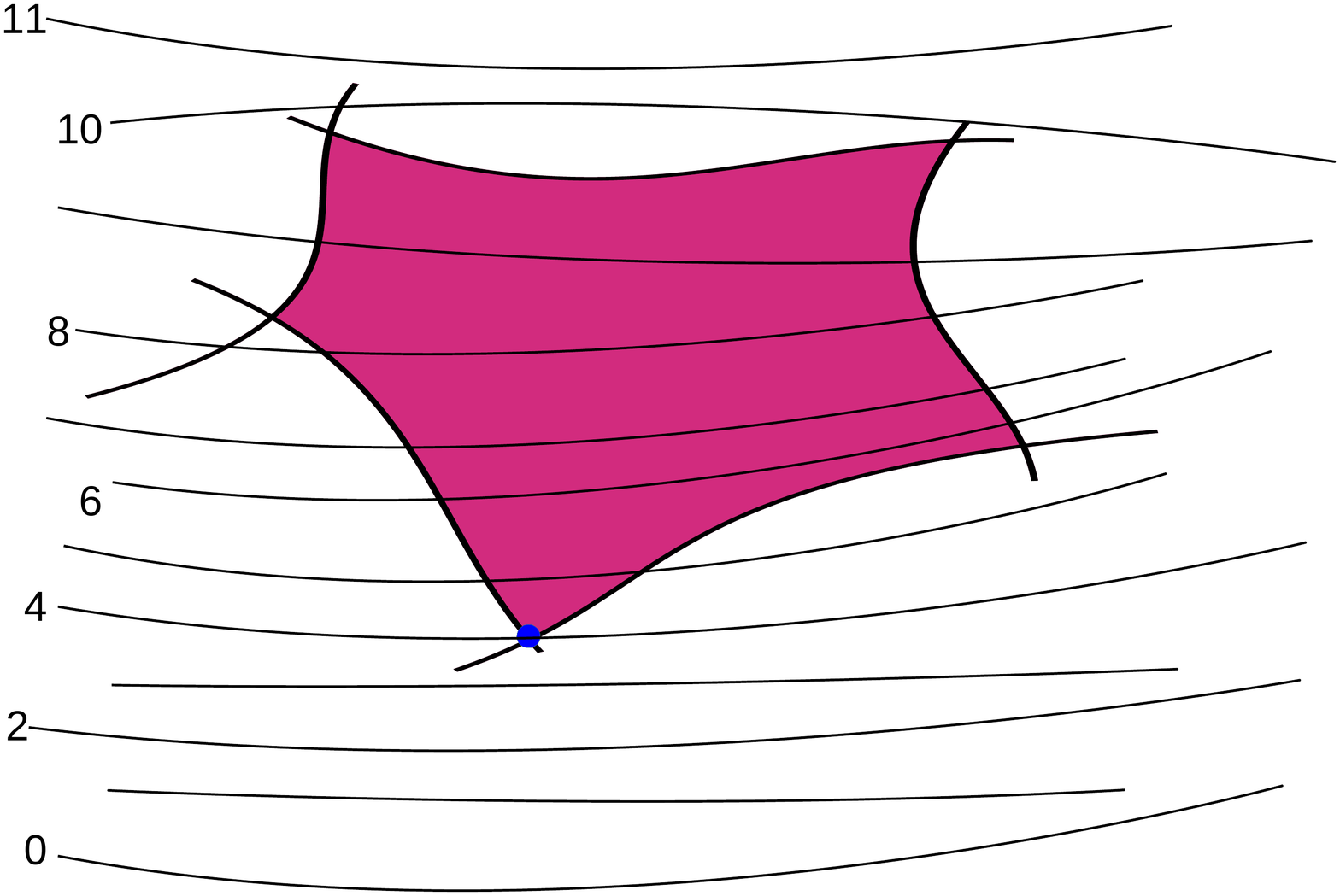}
\caption{Contour curves.}
\end{figure}

{\bf Very often Minimum of Function occurs in Corners}

The crucial point is, that - especially
in a high dimensional coupling-constant
and parameter space - the minimum will
very often fall in a {\bf corner} (where
several sides cross) of the
polyhedron-like region with curved sides
(violet).

But the ``sides'' correspond to different
vacua having zero energy density. So a
corner corresponds to several vacua all
having zero energy density $\rightarrow$
``Multiple Point Principle''!

{\bf 3. Theoretical Reason; Our Bennett's and Mine Original Explanation}
\cite{7mp,8mp,9mp,DonThesis}

One assumes, that some {\bf extensive
quantities / commodities} i.e. some
integrals over space time of say
fields raised to some powers etc. - say
Higgs field squared - are {\bf fixed}
by ``God''/ some law, rather than as I
think we would usually think, it is the
couplings themselves that are selected by
``God''.

To really obtain the Multiple Point Principle as we
wanted you must fix some integrals over the four dimensional
space time to ``God given values''. Mathematically, however,
what we did was very analogous to what one does
for a three dimensional system in working with micro-canonical
ensembles, when one e.g. fixes the energy, the volume, and
the number of moles of say water. Then {\em without
specifying these extensive quantities very accurately,
one can get that the intensive quantities temperature and
pressure gets fine tuned to the triple point}, see figure 6.
In the bottle with water, ice and vapour one has actually
fixed the amount of mols of water molecules, the volume
and the total energy of the system inside the bottle.
That is to say extensive quantities were fixed, but not
to any very special values. The pressure and temperature,
however, come out with the very special triple
point values. The name for our principle is derived in
analogy with this notation ``triple point'' for the
(pressure, temperature) combination appearing at such
a meeting of several (here three) phases.

\begin{figure}
\centering
\includegraphics[scale=0.5]{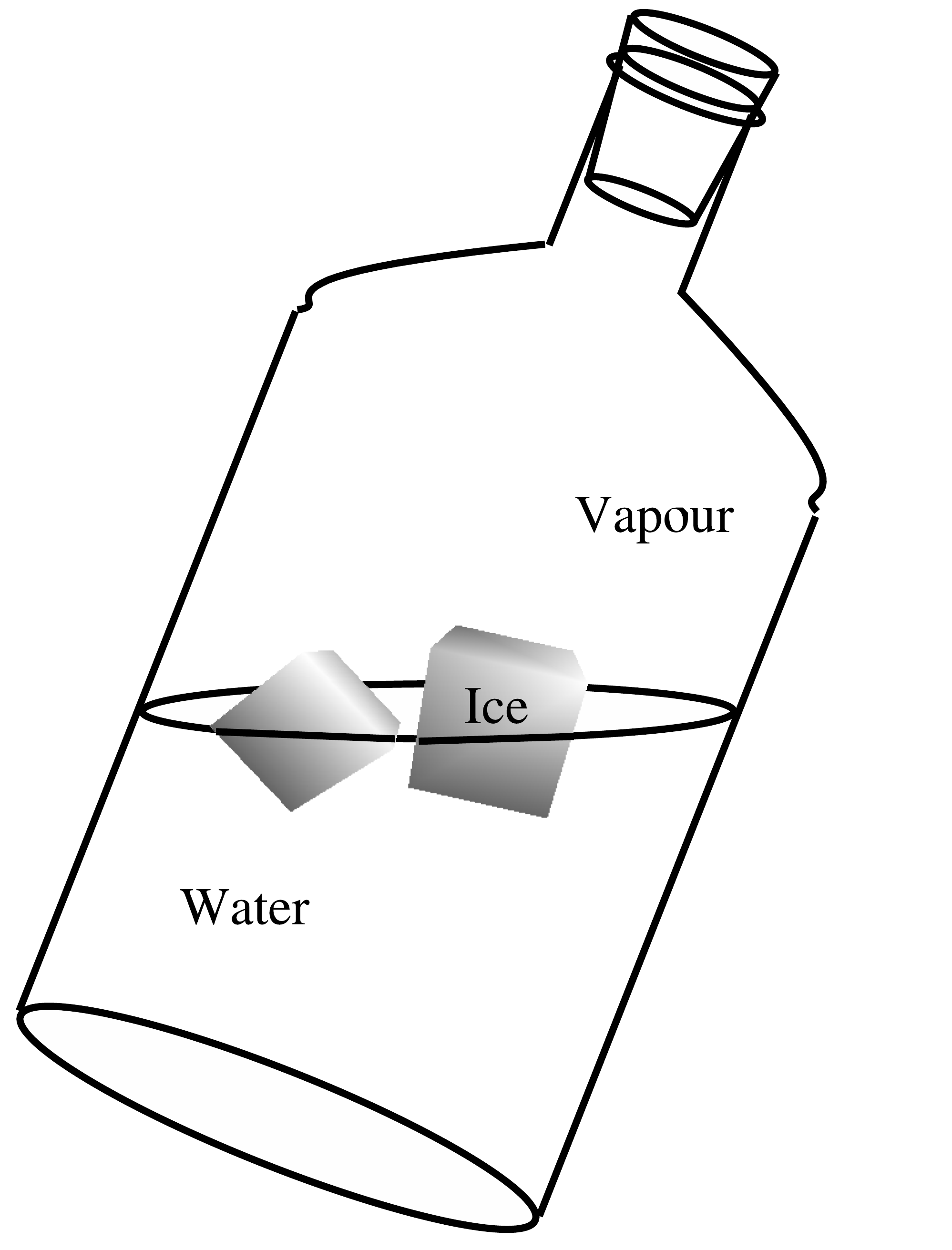}
\caption{Vapour, water and ice equilibrium in a bottle.}
\end{figure}

One should take this analogy of the method for deriving
the Multiple Point Principle with the often found slush
having to have a fixed temperature as very encouraging!
When nature naturally provides slush and just $0^0$ Celsius
for us, it might also make the MPP-situation for the
couplings.

\subsection{Model Reasons}

It should be stressed that in specific model-pictures
one can also derive the ``Multiple (Criticality) Point
Principle'' supposedly though it is only possible in
models allowing the coupling constants to depend on
the {\bf future} too. Ninomiya and one of us (HBN)
``derived'' it in the imaginary action type of theory,
and somewhat similarly in a nonlocal theory by Stillits
\cite{deriving, Stillits} and it was done in babyuniverse
theory, see works by Kawana et al.  \cite{Kawana1,
Kawana3, Kawana4, Kawana5}.
%etc.   \cite{Yamaguchi}.

\section{The Difficulty of the Bound State}

When we - as we now do - want to check
if the ``multiple point principle'' is a
true/valid law of nature, we have the
difficulty that an important role
is played by a
bound
state with which the vacuum, which we call
``condensate vacuum'', is filled.

{\bf
Fundamentally one cannot
calculate completely perturbatively, when
one calculates on a bound state!}

Especially when, as we expect, the bound state is
very strongly bound, the calculation will
involve a non-perturbative calculation, while in
quantum field theory the only truly working method is
perturbation theory. So we are advised to make doubtful
approximations using less reliable techniques such as
bag models or mean field theories, or we have to be
extremely lucky in that something can be actually calculated
accurately. Really one should in the long run make
some very hard numerical calculations by computer, but
so far we have ourselves only made very poor computations
by hand waving approximations.
\begin{figure}
\centering
\includegraphics[scale=0.52]{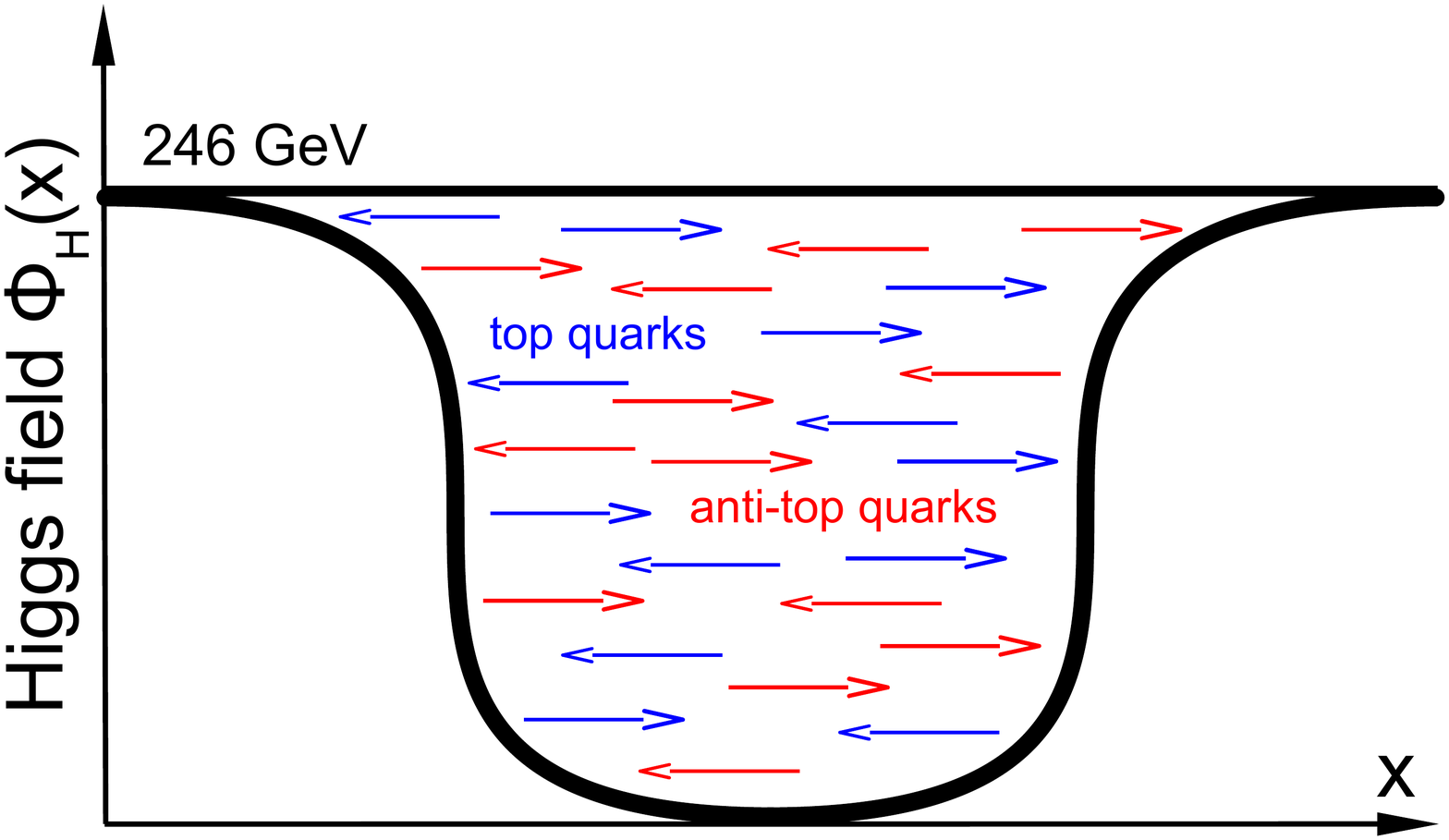}
\caption{Bag in Higgs field filled with top and anti-tops.}
\end{figure}

{\bf Lucky Overdetermined Situation}

Luckily we are with the ``multiple point
principle''(to be tested) in the very
good situation calculationally, that in
addition to knowing already from experiment
all the parameters of the Standard Model
today, we have the three extra
equations, if MPP is assumed. So we
can use this information to help us
through the caculational problems with
the bound state.

{\bf Checking MPP by Calculating Mass of
the Bound State in Several Ways}

Since the calculation is non-perturbative
and thus difficult so that we can only
very crudely
%either difficult or very crude only that we can
compute say the mass of the bound
state, it is suggestive to take it as a
parameter. Then we may formulate testing
the ``multiple point principle'' as
evaluating, by {\em different} assumptions
inside MPP, the mass of the bound state
several times and see if we shall get the
same mass each time we calculate (by using different
information from MPP).

{\bf Really Check Bound State Mass
Obtained from Degeneracy of Vacuum Pairs}

Our technique - today - is to
estimate/calculate the value of the mass
of the bound state of 6 top and 6 anti top
quarks speculated to exist in our
picture/model. (Since we have two
relative energy density predictions
- ignoring the absolute smallness
of the cosmological constant - we get two
such bound state mass fits.) In addition
we can seek to obtain the bound state
mass by building a bag-model-like ansatz
for the bound state and estimates its mass.
Thus we get, by using our multiple point
principle, two a priori different mass
predictions for the bound state,
in addition to the bag model estimate
of the mass as being that of a bound state.

{\bf Our Three Bound State Mass Fits:}

\begin{itemize}
\item {\bf High field fit}: We fit to get
a tiny correction to the Higgs mass
relative to the running self-coupling so
as to ensure the MPP - requirement that
the ``present vacuum'' be
degenerate with the ``high field vacuum'':
Fitting mass $m_{from \; high \; field \; fit}
\approx$ 700 GeV to 800 GeV.

\item{\bf Condensate vacuum fit}: We fit the
mass to be such that the binding between the bound
states in a region filled with such
particles in the lowest energy density state just
gets zero/same energy density as in the
present vacuum.  With a simple but
accidentally almost true assumption,
discussed in section 7 below,
we fit the mass  to $m_{from \; condensate  \; fit}$ $ \approx 4 m_t$ = 692 GeV
$\pm$ 100 GeV, say.
\item{\bf Ansatz calculation}: We make a
bag-model- like crude ansatz for the bound state
of the 6 top + 6 anti top quarks and
seek the minimum energy/mass by varying
bag radius $R$. With  very crude
inclusion of
various corrections, we reach the
mass estimate $m_{bag-model}\approx
5 m_t$  = 865 GeV $\pm$ 200 GeV, say.
\end{itemize}

\section{Plan}

{\bf Introductionary sections:}
\begin{itemize}
\item{{\color{blue}Heading}}: F(750), We miss you!
\item{{\color{blue}Introduction}}: New Natural Law of
Nature, and Bound State.
\item{{\color{blue}MPP}}: The New law,
``{\color{blue}M}ultiple
{\color{blue}P}oint (Criticality) {\color{blue}P}rinciple''.
\item{{\color{blue}Bound}} Bound state of 6 top and 6 anti top quarks.
%\item{Plan}: Plan of the talk.
\item{{\color{blue} Reasons}}: Attempts to explain why MPP.
\item{{\color{blue}Plan}}: Plan of the talk.
\end{itemize}

{\bf Three Calculations:}
\begin{itemize}
\item{{\color{blue}High}}: The mass of the bound state
that could arrange the stability of
our  vacuum to be just borderline
stable w.r.t. the Higgs field.
\item{{\color{blue}Condensate}}: The mass of the bound
state making the ``condensate vacuum''
degenerate in energy density to the
``present vacuum''.
\item{{\color{blue}Bag}}: Bag model estimation of the
mass of the bound state.
\end{itemize}

{\bf Important Application plus Conclusion:}
\begin{itemize}
\item{{\color{blue}Hierachy}}: Our multiple point principle enforces
the weak scale of energy compared to the one of the
high higgs field $\sim$ Planck scale.
\item{{\color{blue}Conclusion}}: Telling that you
should now believe our MPP law of nature!
\end{itemize}

\section{High}
{\bf Small Correction by Laperashvili, Das, and me to ``High field Vacuum'' Energy Density }\cite{LNvacuumstability}

From De Grassi et al.'s calculation\cite{Deg} of the
effective Higgs field potential
$V_{eff}(\phi_H)$ there is a high scale minimum in
this potential, but it goes {\bf slightly}
 under
$0$ so that the present vacuum is
{\bf unstable} for the experimental
Higgs mass $125.09 \pm 0.24$, while the
value that would have made the second
minimum just degenerate with the present
vacuum energy density would be rather
$m_H|_{from  \; MPP \; De \; Grassi...} = 129.4$
GeV.

We have claimed that with a bound state e.g.
with the mass 750 GeV we would get
corrected the De Grassi et al. calculation
so as to be consistent with exact MPP.

Basically we claim: The leading diagrams
treating the bound state as an
``elementary particle'' (i.e. with formal
Feynman rules) though modified by
including estimated form factors, we can
fit the mass of the bound state so that
the diagrams just cancel the instability
 and make the energy density of the
high Higgs field minimum exactly zero,
$V_{eff}(\phi_H \sim 10^{18}\ \hbox{GeV}) =0$.
That works for a mass $\sim$ 700 to 800 GeV.

It should be had in mind that, with the experimentally
suggested numbers, the Higgs mass squared term in the
effective potential for the Higgs field becomes negligible as
soon as we go to field scales high above the weak
scale. Thus for the interesting scale of the
``high field vacuum'' the effective potential
is practically given by the fourth order self coupling
term with coefficient $\lambda_{eff}(\phi_H)$. Indeed to
first approximation one can take the running coupling
for the $\lambda$ taken at the value of the Higgs
field $\phi_H$. If, as we predict\cite{9mp} by MPP, the effective
potential shall just touch zero at the high field vacuum scale,
then it means that approximately also the self coupling
should just touch zero there.

\begin{figure}
\centering
\includegraphics[scale=0.6]{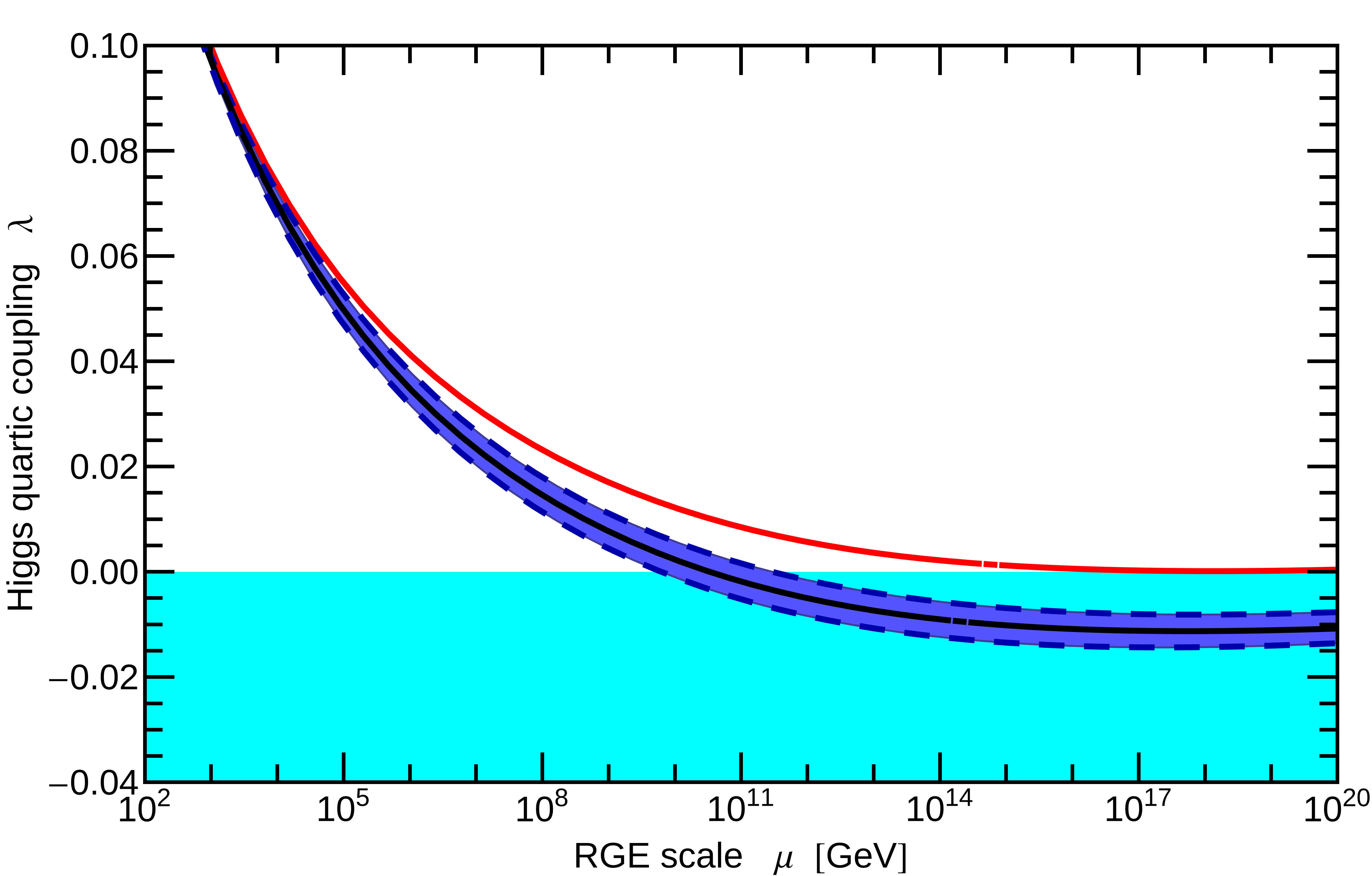}
\caption{Higgs quartic coupling $\lambda$ RGE running.}
\end{figure}

The running of the Higgs self-coupling is somewhat sensitive
to the top-Yukawa coupling $g_t(t)$ and thus to the top mass
$m_t$ from which of course $g_t$ is determined experimentally.
It is therefore practical to present our prediction of the
Higgs mass as a function of the top-mass. One shall then
have in mind that our MPP prediction is that the ``present
vacuum'' is just barely stable - it has namely just the
same energy density as the ``high Higgs field one'' - and
thus our prediction is that the
combination of Higgs and top masses shall be on the line
separating the stability and the meta-stability of the
present vacuum on the following plot figure 9.
\begin{figure}
\centering
\includegraphics[scale=0.6]{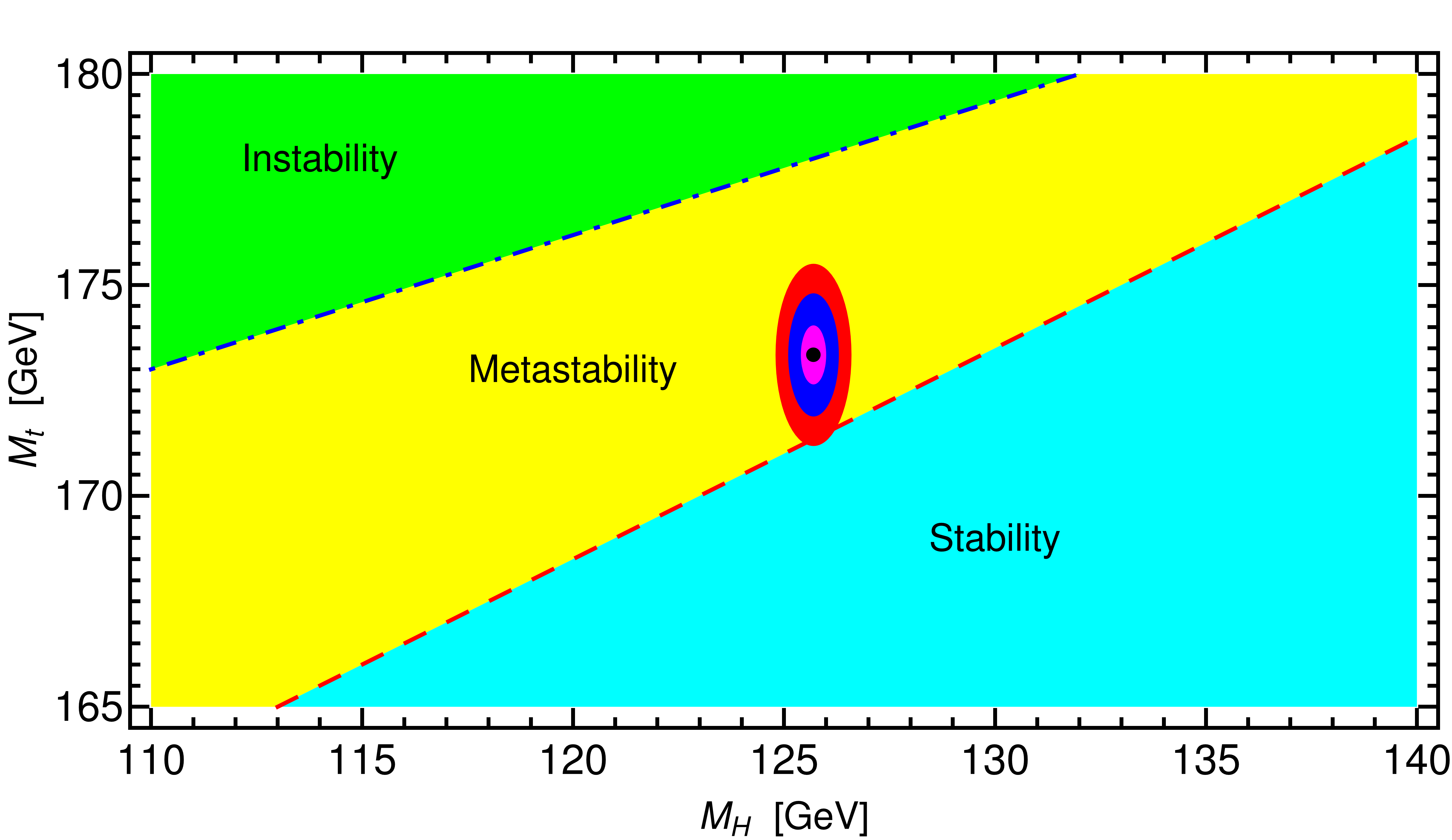}
\caption{Instability, metastability and stability region.}
\end{figure}

\begin{figure}
\centering
\includegraphics[scale=0.9]{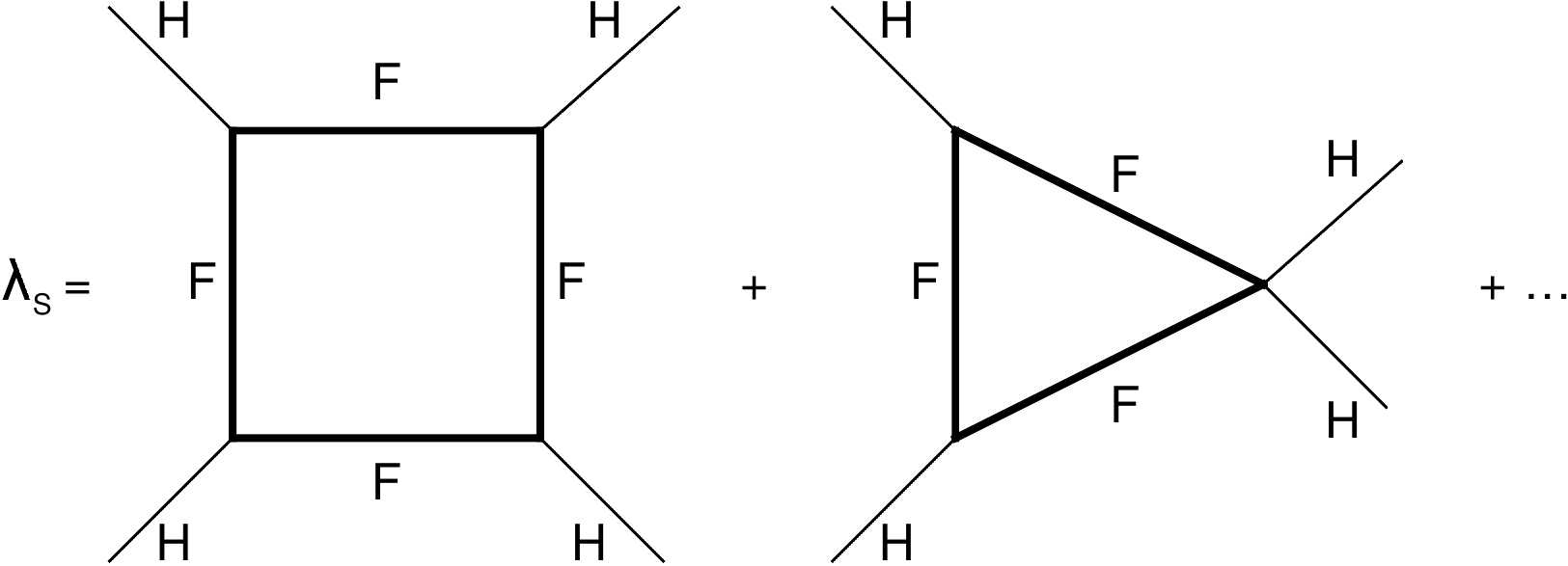}
\caption{Correction to Higgs-self coupling.}
\end{figure}

{\bf The Small Instability,
Negative Self-coupling at the High
field Minimum:}

Extrapolated using DeGrassi et al.\cite{Deg} without
our correction, one gets the following
value of the running self coupling
$\lambda_{run}(10^{18}\ \hbox{GeV})$:
\begin{equation}
\lambda(\phi_{\hbox{``high field''}})
= -0.01 \pm 0.002.
\end{equation}
at the high field scale
$\phi_{\hbox{``high field''}}$.
However,
 a value very accurately zero is
required by the Multiple Point Principle (=MPP).

Since the bound state F
%identified with
%F(750)
is an extended object we must
include a form factor, when using it
in Feynman diagrams.
A priori it is of course a very doubtful approximation
to describe a bound state like ours as if it were a fundamental
particle and then only improve the calculation by
introducing form factors taking the extended structure of the
bound state into account. In the light of the fact that our
bound state is expected to be very strongly bound - the
mass being about 1/3 of the collective mass of the
constituents - there is a better hope that such an
as if fundamental particle approximation could be
more justified. But it is still one of the very
doubtful approximations in our works with this
bound state.

Now to obtain form factors you need the radius of the
bound state.

Defining a quantity $b$ denoting the
radius of the bound state measured with the
top quark Compton wave length $1/m_t$ as
unit by:
\begin{eqnarray}
\langle\vec{r}^2\rangle& =& 3r_0^2,\\
r_0 &=& \frac{b}{m_t},
\end{eqnarray}
 we obtain a theoretical estimate
\begin{equation}
b=\sqrt{\frac{\langle\vec{r}^2\rangle}{3}} m_t \approx 2.34. \label{9w}
\end{equation}

{\bf Approximating the Bound State as if it were
an Elementary Particle, since so
Strongly Bound}

The dominant diagram/correction - the
first (box) diagram on
the figure just
above - is
$$ \lambda_S \approx
\frac{1}{\pi^2}\left(\frac{6g_t}{b}\times
\frac{m_t}{m_S}\right)^4$$
where we have the estimated or measured
values
$$ g_t =0.935 ;\;  m_t = 173\ \hbox{GeV};\;
b\approx 2.34.$$
%\hbox{or} 2.43$$

Using the after all rather small deviation
from perfect MPP
$$\lambda_{\hbox{high field}} = -0.01 \pm 0.002$$  and requiring it to be cancelled by
the correction from the bound state we
get the requirement
\begin{equation}
\lambda_S = \frac{1}{\pi^2}\left (
\frac{6g_t}{b}\times\frac{m_t}{m_{F}}\right )^4 \times
( \sim 2) \approx 0.01 \pm 0.002,
\end{equation}
where $g_t = .935$, $m_t=173\ \hbox{GeV}$,
$b\approx 2.34$ and the factor
``$(\sim 2)$'' were taken in to
approximate some neglected diagrams.

If a further study should show that
the next diagrams add up to roughly as
much as the first one, we should include the
factor $\sim 2$ to take into account
the neglected Feynman diagrams correcting
the Higgs self coupling.

The solution w.r.t. the mass of the
bound state $m_{F}$ gives
\begin{eqnarray}
m_{F} &\approx& \frac{6g_tm_t}{b}
\left (
\frac{\sim 2}{\pi^2\times 0.01\pm 0.002}\right)
^{1/4} \nonumber\\
\approx 2.31\times173\ \hbox{GeV}
\times2.1& =&4.9 \times173\ \hbox{GeV} =850\ \hbox{GeV} \pm 20 \%
\nonumber\\
\hbox{or without the $\sim 2$:}\;&&\\
m_{F}= 2.31 \times173\ \hbox{GeV} \times 1.8&=&
4.1 \times173\ \hbox{GeV} =710\ \hbox{GeV} \pm 20 \%
\nonumber
\end{eqnarray}

{\bf Three Agreeing Fits of the Bound
State Mass:}

Let us here tell that we soon shall present
more bound state calculations than the one just
presented for which we give two
result values, depending on whether the next term
in the Feynman diagram expansion will
be important or not. Only the very first
- box diagram  - term  gives what we call
``without $\sim 2$'' factor.

The summary, before having presented details
of the other crude bound state calculations, is:
\begin{eqnarray}
m_{F}(\hbox{from ``high field vacuum''})
&\approx & 850\ \hbox{GeV}\pm 30 \%
\ \hbox{with $\sim 2$}\\
m_{F}(\hbox{from ``high field vacuum''})
&\approx & 710\ \hbox{GeV}\pm 30 \%
\ \hbox{without $\sim 2$}\\
m_{F}(\hbox{``condensate vac.''})&\approx &
692\ \hbox{GeV}\pm 40 \% \\
m_{F}(\hbox{``bag estimate''})&\approx
& 5 m_t = 865\ \hbox{GeV}\ \hbox{(very uncertain)}.
\end{eqnarray}

The agreement of the value ``692 GeV''
with the estimate(s) from the
completely different vacuum with
the high Higgs field ``850 GeV''
or ``710 GeV'' and with the mass by
estimating how strong the top and anti
tops can bind $m_{F}(\hbox{``bag estimate''})
\approx 865\ \hbox{GeV}$   is
encouraging and provides support for our
``Multiple Point Principle''!

\section{Condensate}
{\bf Fitting Bound State Mass to the ``Condensate Vacuum''
having Same Energy Density as the ``Present'' one.}\cite{mass,
Tunguska}

For calculational purposes we approximate
the ``condensate vacuum'' with a crystal
(actually it should at least be a fluid, but
that may not matter so much for our crude
energy density estimate) made from the
bound states sitting each with 4 neighbors,
the tops and anti tops of which are in
approximate main quantum number n=2 states seen
from the bound state considered.

The MPP-requirement
may be  written
\begin{eqnarray}
0 &=& m_{F} - ``\hbox{binding
per F}''\\
  &=& m_{F}-
\frac{\#\hbox{neighbors}}{2}
\times ``\hbox{binding to neighbor F}''\\
&\approx& m_S - \frac{4}{2} \times
``\hbox{binding
of F in n=2 arround another
F}''\\
&\approx& m_{F} - \frac{4}{2}\times
``\hbox{binding of F}''\times
\frac{1/2^2}{1/1^2}\\
&=& m_{F}-\frac{1}{2}\times
``\hbox{binding of F}''\\
&=& m_{F} -\frac{1}{2}\times(12m_t-m_{F})\\
&=& \frac{3}{2}m_{F} - 6 m_t
\label{zerocond}
\end{eqnarray}

We here follow an appendix of
our earlier work\cite{Tunguska} and
assume
that the structure of the condensate can
be approximated as being a diamond lattice
structure, so that there are just
$\#\hbox{neighbors} = 4$ neighbors, i.e. other
F-particles
surrounding each one of them in the
lattice. When we count all the binding
energy per F present in the
condensate
$``\hbox{binding per F}''$ as being
the
number of neighbors $\#\hbox{neighbors}
$ times   the binding
of one F to its neighbor
$``\hbox{binding to neighbor F}''$,
we {\em double count}, because we
count the same binding from both the
one F and from the other one it
binds to.

We made then the approximation
%that we can effectively consider it,
that the neighboring top quarks and
anti top quarks contained in an F
neighboring to another one are {\em in
effect
in the n=2 orbit} of the latter. Thus we
can take the binding energy of a
neighboring F to a given one
$``\hbox{binding to neighbor F}''$
to be  as if the tops and anti tops
were in an n=2 orbit or some
superposition thereof. Thus the binding
of the neighbors occur with binding energy
``binding
of F in n=2 around another F''.

As long as we can take the effective
Higgs mass for the two lowest orbits
n = 1 and 2 to be zero, we can count that
the binding energy, for top say, in the
n=2 orbit is just one quarter of that
in the n=1 orbit, provided we can use the
same potential of the form $\propto 1/r$.
%But now that is what our above discussion
This result is ensured by an
``accidental cancellation'' discussed in
an earlier article\cite{4nbs}, which we now briefly
explain.

When one considers a top quark say circulating around a bunch of
eleven top or anti top quarks, the attraction is only
about half as strong as it would be if these 11 particles
were all concentrated in a single point/a center. This is because
on the average half of the 11 particles are further away from the center
than the considered particle. Thus for such a group
of particles binding to each other, one should indeed use an effective coupling
strength, say $g_t^2$, which is only half as strong as what we get formally.
In the article\cite{4nbs}, in which we calculate the mass of the bound state needed
for making the F condensate be degenerate with the present vacuum,
we found that this correction by the factor 1/2 in the strength of
attraction happens to approximately cancel a couple of
other corrections. So that very nicely we could ignore this correction
and take it that we had thereby even corrected for the gluon
exchange attraction, as well as for the effect of exchange
of W's and Z's (or what we call eaten Higgses, since it is mainly
the longitudinal Z or W that matters).

In this way the calculation simplifies appreciably and
even for an F-particle, which consists
of tops and anti tops, the ratio of the
binding energies in the two lowest orbits
should be $1/2^2 = 1/4$.

From the last step in (\ref{zerocond})
we easily derive of course that
\begin{equation}
m_{F} = \frac{2}{3}\times6m_t = 4m_t
= 4\times173\ \hbox{GeV}
= 692\ \hbox{GeV agreeing well with }
710\ \hbox{GeV  or } 850\ \hbox{GeV!}
\end{equation}

\section{Bag}
{\bf Estimating How 6 Top + 6 Anti top
Bind}

We have made an estimate of
the mass of the bound state of
6 top + 6 anti top quarks,
to be crudely
\begin{equation}
 m_{bound \; state} =
 % 787
 875\ \hbox{GeV}
 \hbox{ or }
 % 761
 792\ \hbox{GeV} \pm \hbox{say 40 \%}.
 \end{equation}
Our method mainly used a
bag-model estimation, in which
the bag meant a region where the
Higgs field was reduced to
$\sim 0$.

Such a region, where the Higgs field expectation value
goes approximately to zero, we expect to occur for sufficiently
strong interaction of the top and anti top
quarks with the Higgs.

Assuming that such a bag of very small Higgs field value
becomes significant we estimated, using some further adjustments
of how much the top or anti top quarks may tunnel out of
the bag, how to obtain the minimal energy ansatz for the
bag system with the 12 top or anti tops inside. We managed
to get the mass of the bound state down to $ \sim 7 m_t$
$ \approx  1216\ \hbox{GeV}$. Then, however, including further
corrections such as the exchange of W's and Z's (or, as we call
them, eaten Higgses) and gluons, crude corrections of the bag
calculation rather give
\begin{eqnarray}
M_{corrected \; eaten + gluons} =
%0.247
%0.320 m_t
%+ 4.74 m_t =
%4.99
5.06 m_t=
%787
875\ \hbox{GeV}.
\end{eqnarray}

%Imagining that the top-Yukawa-coupling
%$g_t$ were gradually screwed up, the Higgs
%field inside an ansatz bound state of 6 tops + 6 anti tops,
%at say the typical distance of the quarks themselves from the center,
%would gradually be lowered compared to the usual vacuum expectation value.

\section{Hierarchy}
{\bf ``Solve'' Hierarchy problem}\cite{1nbs,2nbs}

Having made a fine-tuning
theory/model/rule we have at least the
chance to have our fine-tuning theory
MPP give the experimentally observed order
of magnitude for the Higgs mass say. And
indeed we predict the right order for the
logarithm of the scale range over which
$\mu$ has to run to get the running
top-yukawa-coupling $g_{t \; run}(\mu)$ go
from $0.4$ at the $10^{18}$ GeV to the
0.935 needed at the weak scale, from the
requirement of the ``condensate vacuum''
being degenerate with the ``present one''.

\begin{figure}
\centering
\includegraphics[scale=0.5]{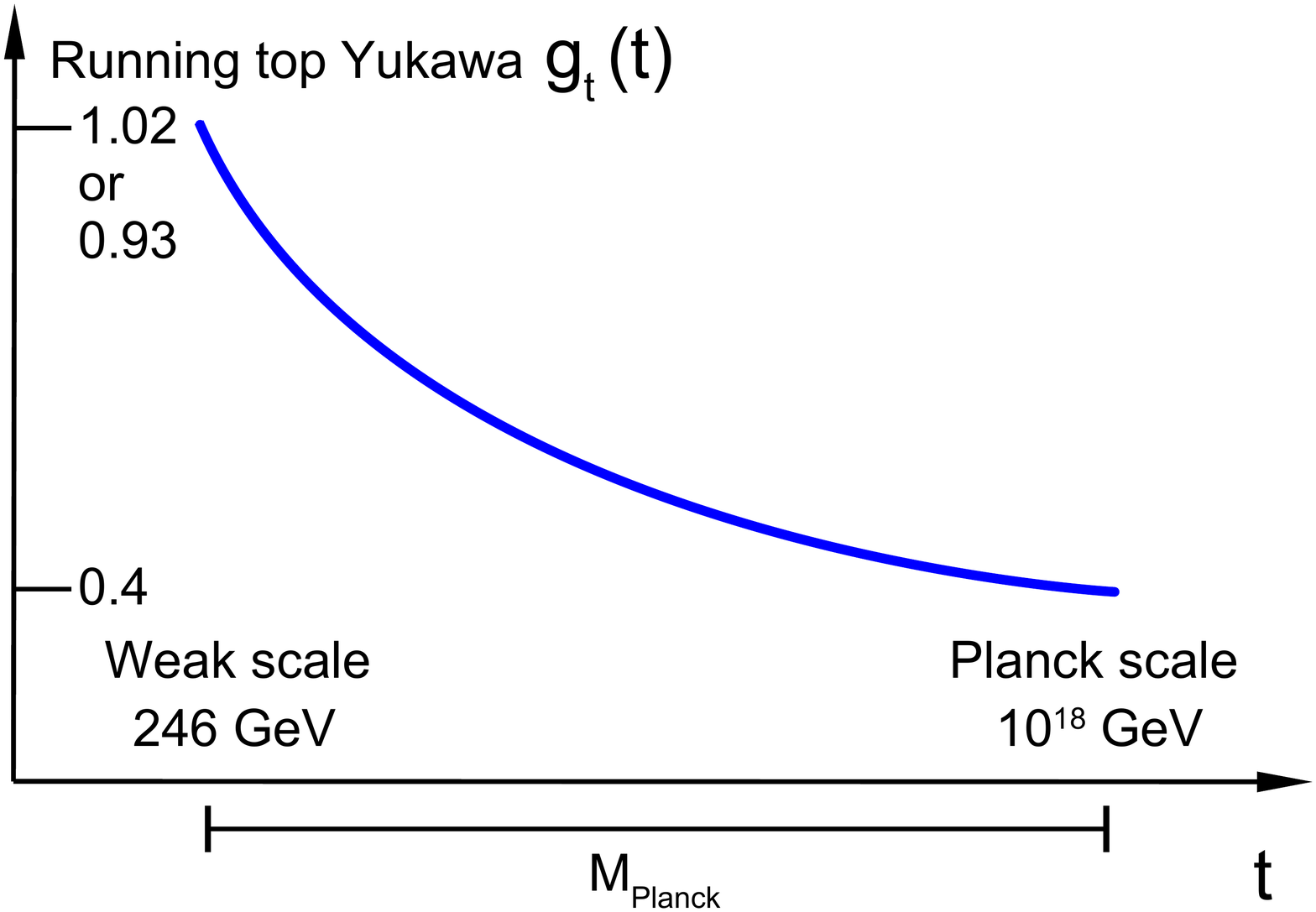}
\caption{Top Yukawa coupling RGE running.}
\end{figure}

It should be stressed that, on the figure illustrating
the running top Yukawa coupling $g_t(t)$ as a function
of the scale(parameter) $t$, our multiple point principle
predicts the value of this top Yukawa {\em both} at
about the Planck scale $10^{18}$ GeV and at the weak scale.
We can thus in principle use the knowledge of the rate
of running, the beta-function, to predict from MPP the
scale of the weak interactions. Our success is that this
weak scale becomes very much smaller than the Planck scale,
and in fact it even comes out very well! In that sense
we did not really ``solve'' the hierarchy problem, because
it is still a fine tuning, but we have by MPP provided
a theory for the fine tuning!

\section{Conclusion}
{\bf Conclusion}

The remarkable {\bf coincidence}, that
our three mass estimations coincide provides
evidence in favour of the {\bf truth
of our model, with the Multiple Point
Principle and the bound state}!

{\bf Some Achievements of our Model MPP
and Strongly Bound State of 6 Top + 6 Anti top quarks (in pure Standard Model)}
\vspace{-2mm}

I must mention the following achievements
most of which I did not have much time for:

\begin{itemize}
\item{{\color{red} Hierarchy problem}}:
The fine-tuning caused by our MPP
requirements combined with the assumption,
that the Higgs field in the ``high field
vacuum'' is of the order of the Planck
scale (or only a bit under) leads to the
scale problem being solved, in the sense
that the Higgs mass and weak scale get
fixed to be {\em exponentially} much
lower than the Planck scale, and that in
fact we get very close to the right size for
the logarithm.

\item{{\color{red}$g_t$}}: Froggatt and I
estimated the value of the
top-Yukawa-coupling $g_t$ needed for MPP,
in the sense that it represents a phase
transition value between the
 ``condensate vacuum''
and the ``present vacuum''. We found the
phase transition value
$g_{t \; phase \; transition}= 1.02 \pm 14\%$,
agreeing  with  experiment $g_{t \; exp}
= 0.93_5$.
\item{{\color{red} Stability}}: Explaining
that the Higgs mass
just puts our vacuum on the borderline
of being meta-stable\cite{9mp}.
\item{{\color{red} Correction
to Stability}}: ...even very accurately,
if we
take seriously the very small correction
due to the bound state by Laperashvili,
Das, and myself.

\item The ``condensate vacuum'' can be
used for a model for dark matter  as
pearl size balls of the ``condensate
vacuum'' surrounded necessarily by a skin
- the transition surface - that is then
pumped up by ordinary matter, carbon say,
to a pressure of the order of that in
a white dwarf star. Such pearls may be
useful for
\begin{itemize}
\item Dark matter \cite{Tunguska,Dark1,Dark2}.
\item making supernovae explode so as to
throw sufficient material out so that we
can observe them \cite{Supernova}.
\item Explaining the two bursts of
neutrinos observed with $\sim$ 5 hours
time difference in SN1989A in the Big
Magellanic Cloud \cite{Supernova}.
\item helps r-process fit ?
\item explain ratio of dark to normal
matter being of order 6 \cite{Dark1,Dark2}.
\end{itemize}
\end{itemize}

{\bf Achievements of MPP in Extended
Models (i.e. not only Standard Model):}
\begin{itemize}
\item{{\color{red}Value of Cosmological
Constant}}: With Roman Nevzorov we got
values for the CC using the ``same version''
of MPP and an almost supersymmetric vacuum
state.
\item{{\color{red}Number of families }}:
Prior to having formulated MPP we fitted
fine structure constants in an extension
of the Standard Model ``AntiGUT'' in which
each family of fermions has its own set
of gauge bosons. We - including Brene
and Don Bennett and me - {\bf pre}dicted
the number of families, which was not
known yet at that time.
\end{itemize}

{\bf Encouragement for Theoreticians to Calculate More
Accurately This Bound State}

At the end I would stress:

{\bf Since our picture is PURE STANDARD
MODEL, everything can in principle be
CALCULATED!}

So it is only a question of better
techniques - Bethe Salpeter Equation ?\cite{BS} -
or better computers and use of them
- lattice theory with Higgs field on the
lattice ? -  to obtain more solid and
accurate checks of MPP and calculation
of the bound state mass than my crude
estimates.

And this is just a work for the
theoreticians (among the students say).

Then there should pop up some peaks
- like the joke of Pich's (mentioned at this conference
as a joke\cite{650} from CMS) - by themselves,
when the experimentalists make the plots.

If not it would mean that the Standard Model
were not right also nonperturbatively, if our calculations were right.

\section*{Acknowledgement}
Holger Bech Nielsen thanks the Niels Bohr Institute
for allowance to stay as emeritus and for some travel support,
which though seems to have run out before the travel to
Corfu and back were finished.

%\section{...}

\end{document}